\documentstyle[epsfig,12pt]{elsart} 
\begin{document}
\newcommand{\kp}{K^+}
\newcommand{\gk}{\vec{\gamma}\vec{k}}
\newcommand{\gE}{\gamma_0 E_k}
\newcommand{\ppl}{\vec{p}}
\newcommand{\bcm}{\vec{b}^{\star}}
\newcommand{\becm}{\vec{\beta}^{\star}}
\newcommand{\bepl}{\vec{\beta}}
\newcommand{\rcm}{\vec{r}^{\star}}
\newcommand{\rpl}{\vec{r}}
\newcommand{\A}{{$\mathcal A$}}
\newcommand{\wpk}{ \omega_{p-k}}
\newcommand{\Journal}[4]{ #1 {\bf #2} (#4) #3}
\newcommand{\NPA}{Nucl.\ Phys.\ A}
\newcommand{\PLB}{Phys.\ Lett.\ B}
\newcommand{\PRC}{Phys.\ Rev.\ C}
\newcommand{\ZPC}{Z.\ Phys.\ C}
\newcommand{\be}{\begin{equation}}
\newcommand{\ee}{\end{equation}}
%%%%%%%%%%%%%%%%%%%%%%%%%%%%%%%%%%%%%%%%%%%%%%%%%%%%%%%%%%%%%%%%%%
\begin{frontmatter}

\title{Fragment formation in proton induced reactions 
within a BUU transport model}

\author{T. Gaitanos}
\author{H. Lenske}
\author{U. Mosel}

\address{Institut f\"ur Theoretische Physik,
Universit\"at Giessen, D-35392 Giessen, Germany}
\address{email: Theodoros.Gaitanos@theo.physik.uni-giessen.de}
%%%%%%%%%%%%%%%%%%%%%%%%%%%%%%%%%%%%%%%%%%%%%%%%%%%%%%%%%%%%%%%%%%%
\begin{abstract}
The formation of fragments in proton-induced reactions at low 
relativistic energies within a combination of a covariant 
dynamical transport model and a statistical approach is investigated. 
In particular, we discuss in detail the applicability and limitations 
of such a hybrid model by comparing data on fragmentation at 
low relativistic $SIS/GSI$-energies. 
\end{abstract}
%\pacs{25.75.+r}
%******************************************************************
\end{frontmatter}

%\end{quote}}
\date{\today}
%\maketitle
%\setpapersize{A4}
      %    Width of rule between columns.
%%%%%%%%%%%%%%%%%%%%%%%%%%%%%%%%%%%%%%%%%%%%%%%%%%%%%%%%%%%%%%%%%%%
%%%%%%%%%%%%%%%%%%%%%%%%%%%%%%%%%%%%%%%%%%%%%%%%%%%%%%%%%%%%%%%%%%%
%%%%%%%%%%%%%%%%%%%%%%%%%%%%%%%%%%%%%%%%%%%%%%%%%%%%%%%%%%%%%%%%%%%
%           BEGIN OF TEXT                 %
%%%%%%%%%%%%%%%%%%%%%%%%%%%%%%%%%%%%%%%%%%%%%%%%%%%%%%%%%%%%%%%%%%%
%%%%%%%%%%%%%%%%%%%%%%%%%%%%%%%%%%%%%%%%%%%%%%%%%%%%%%%%%%%%%%%%%%%
%%%%%%%%%%%%%%%%%%%%%%%%%%%%%%%%%%%%%%%%%%%%%%%%%%%%%%%%%%%%%%%%%%%

%%%%%%%%%%%%%%%%%%%%%%%%%%%%%%%%%%%%%%%%%%%%%%%%%%%%%%%%%%%%%%%%%%%
\section{Introduction}
%%%%%%%%%%%%%%%%%%%%%%%%%%%%%%%%%%%%%%%%%%%%%%%%%%%%%%%%%%%%%%%%%%%

One of the major aspects in investigating proton induced reactions 
is to better understand the phenomenon of 
fragmentation of a nucleus in a hot fireball-like state. 
Proton induced reactions are the simplest possible way to study such 
phenomena. Also, they have been found to be important for other 
investigations, e.g., for the production of radioactive 
beams \cite{intro1} or to interpret the origin of cosmic rays and 
radionuclides in nuclear astrophysics \cite{intro2}. More recently they have 
gained again experimental \cite{spaladin} interest. It is therefore a challenge 
to study this field of research in detail, in particular in relation to the 
future investigations at the new experimental facility FAIR at GSI.

Proton induced reactions (and also heavy-ion collisions) are usually 
modeled by non-equilibrium transport models, see for a review Refs. 
\cite{Horror,kada,dani}. However, the description of fragment formation is 
a non-trivial task, since transport models do not account for the evolution of 
physical phase space fluctuations. The major difficulty here is the 
implementation of the physical fluctuating part of the collision 
integral and the reduction of numerical 
fluctuations using many test particles per physical nucleon, which however 
would require a large amount of computing resources. Attempts to resolve this 
still open problem have been recently started \cite{colonna}. 

The standard approach of phenomenological coalescence models for 
fragment formation has been found to work astonishingly well in heavy ion collisions, 
as long as one considers only one-body dynamical observables, see \cite{flows}. 
In particular, the coalescence model is usually applied in violent heavy-ion collisions, 
in which a prompt dynamical explosion of a fireball-like system is expected, 
with the formation of light clusters through nucleon coalescence. In this dilute 
matter secondary effects are negligible. However, the dynamical situation in 
proton-induced reactions is different. Compression-expansion effects are here 
only moderate and the fragmentation process happens over a long time scale (compared 
to the short lived explosive dynamics in heavy-ion collisions), which is compatible 
with a statistical description of the process.

The whole dynamical picture in proton-induced reactions is therefore modeled by a 
combination of dynamical and statistical models. 
So far, two types of microscopic approaches have been frequently applied in 
proton-induced reactions: the intranuclear cascade (INC) model 
\cite{INC} and the quantum molecular dynamics (QMD) prescription 
\cite{QMD}, in combination with a statistical multifragmentation model (SMM) 
\cite{SMM}. The SMM model is based on the assumption of an 
equilibrated source and treats its decay statistically. It 
includes also sequential evaporation and fission. 

In this letter we study fragment formation in proton-induced reactions 
at low relativistic energies in the framework of a fully covariant 
coupled-channel transport equation based on the relativistic 
mean-field approach of Quantum-hadro-dynamics \cite{QHD}. 
As a new feature we consider here 
for the first time the formation of fragments describing the initial step 
dynamically by means of a covariant transport model of a 
Boltzmann-Uehling-Uhlenbeck (BUU) type, followed by a statistical 
formation process of fragments in terms of the SMM model \cite{SMM}. 
We compare the theoretical results with a broad selection of experimental data. 

%%%%%%%%%%%%%%%%%%%%%%%%%%%%%%%%%%%%%%%%%%%%%%%%%%%%%%%%%%%%%%%%%%%
\section{Theoretical description of proton-induced reactions}
%%%%%%%%%%%%%%%%%%%%%%%%%%%%%%%%%%%%%%%%%%%%%%%%%%%%%%%%%%%%%%%%%%%

The theoretical description of hadron-nucleus (and also 
heavy-ion reactions) is based on the semiclassical kinetic theory 
of statistical mechanics, 
i.e. the Boltzmann Equation with the Uehling-Uhlenbeck modification of 
the collision integral \cite{kada,dani}. The covariant analog of this equation
is the Relativistic Boltzmann-Uehling-Uhlenbeck (RBUU) equation
\cite{Ko,giessen}
%%%%%%%%%%%%%%%
\begin{equation}
\left[
p^{*\mu} \partial_{\mu}^{x} + \left( p^{*}_{\nu} F^{\mu\nu}
+ M^{*} \partial_{x}^{\mu} M^{*}  \right)
\partial_{\mu}^{p^{*}}
\right] f(x,p^{*}) = {\cal I}_{coll} 
\quad ,
\label{rbuu}
\end{equation}
%%%%%%%%%%%%%%%
where $f(x,p^{*})$ is the single particle distribution function for the 
hadrons. 
The dynamics of the drift term, i.e. the lhs of eq.(\ref{rbuu}), is
determined by the mean field, which does not explicity depend on 
momentum. Here the attractive scalar field $\Sigma_s$ enters via 
the effective mass $M^{*}=M-\Sigma_{s}$
and the repulsive vector field $\Sigma_\mu$ via the
kinetic momenta $p^{*}_{\mu}=p_{\mu}-\Sigma_{\mu}$ and via the field tensor
$F^{\mu\nu} = \partial^\mu \Sigma^\nu -\partial^\nu \Sigma^\mu$.
The dynamical description according to Eq.(\ref{rbuu}) involves the 
propagation of hadrons in the nuclear medium, which are coupled through the 
common mean-field and $2$-body collisions. The exact solution of the set of the 
coupled transport equations is not possible. Here we use the standard 
test-particle method for the numerical treatment of the Vlasov part. The 
collision integral, i.e. the rhs of  eq.(\ref{rbuu}), is modeled by a 
parallel-ensemble Monte-Carlo algorithm. 

The results presented here are based on Eq. (\ref{rbuu}) in a new 
version of Ref. \cite{GiBUU}, as realized in the Giessen-BUU (GiBUU) 
transport model, presented in \cite{GiBUU,larionov}, where also the 
properties of the relativistic mean-field, cross sections and the 
collision integral are discussed.

Furthermore, we note that the results presented here do not differ essentially 
from those performed with non-relativistic prescriptions \cite{anna}, as will 
be shown below. This is expected, since the achieved energies of the fragmenting 
sources are smaller than the rest energy. However, a fully covariant description 
is advantageous for several reasons. First of all, dynamics and kinematics are 
described on a common level, since the transport equations are formulated in a 
covariant manner. Apart from that, the relativistic mean-field accounts by definition 
for higher order momentum dependent effects which would be missed in a non-relativistic 
approach. The relativistic formulation is also of advantage for other dynamical 
situations, e.g. heavy-ion collisions at high relativistic energies by studying 
the formation of hypernuclei. For these reasons we have used a fully covariant approach, 
which will be applied to the more complex dynamics of relativistic heavy-ion collisions 
in the future.

The non-linear Walecka model has been 
adopted for the relativistic mean-field potential. This 
model gives reasonable values for the compression modulus ($K=200$ MeV) 
and the saturation properties of nuclear matter \cite{larionov,gaitanos}. 
In this first study of multifragmentation (see below) we use its 
standard version accounting only for the iso-scalar part of the 
hadronic EoS ($\sigma,~\omega$ classical bosonic fields). 
The exchange of iso-vector bosons ($\rho,~\delta$) is neglected in the 
mean-field baryon potential. At present, we are interested mainly in 
understanding the fragmentation process in terms of a dynamical description 
accepting the inaccuracies which might occur in the yields for exotic 
nuclei, when isovector interactions are neglected. In order to keep the 
calculations feasible we have to introduce further approximations. We assume 
that all baryons, also hadronic resonances and those with finite strangeness, 
feel the same mean-field. Meson self-energies are not taken into account,  
except for the Coulomb field.

The transition from the fully dynamical (BUU) to the purely 
statistical approach (SMM) is not straightforward, and has to be 
studied carefully. In particular, it involves the time in 
which one has to switch from the dynamical to the statistical 
situation. Furthermore, 
an important feature for proton-induced reactions is the 
numerical stability of ground state nuclei, in particular, the determination of the 
ground state binding energies, within the test particle method. In  
the relativistic non-linear Walecka model the total energy is extracted 
as the space-integral of the $T^{00}$-component of the conserved 
energy-momentum tensor. 
The phase-space distribution function $f(x,p^{*})$ is represented within 
the test particle {\it Ansatz}, in which $f(x,p^{*})$ is discretized in 
terms of test particles of a Gaussian shape 
in coordinate space and a $\delta$-like function in momentum space. 

The transition from the dynamical (BUU) to the statistical (SMM) picture is 
controlled by the onset of local equilibration. We calculate in each 
time step at the center of the nucleus the spatial diagonal components of 
the energy momentum tensor $T^{\mu\nu}$ and define 
the local, e.g. at the center of the source,  anisotropy as 
$Q(x) = \frac{2T^{zz}(x)}{T^{xx}(x)+T^{yy}(x)}$. The onset of local 
equilibration is defined as that time step, in which the anisotropy ratio 
approaches unity ($\pm 10\%$). 
%%%%%%%%%%%%%%%%%%%%%%%%%%%%%%%%%%%%%%%%%%%%%%%%%%%%%%%%%%%%%%%%%%%
\begin{figure}[t]
\unitlength1cm
\begin{picture}(10.,8.0)
\put(1.5,0.){\makebox{\psfig{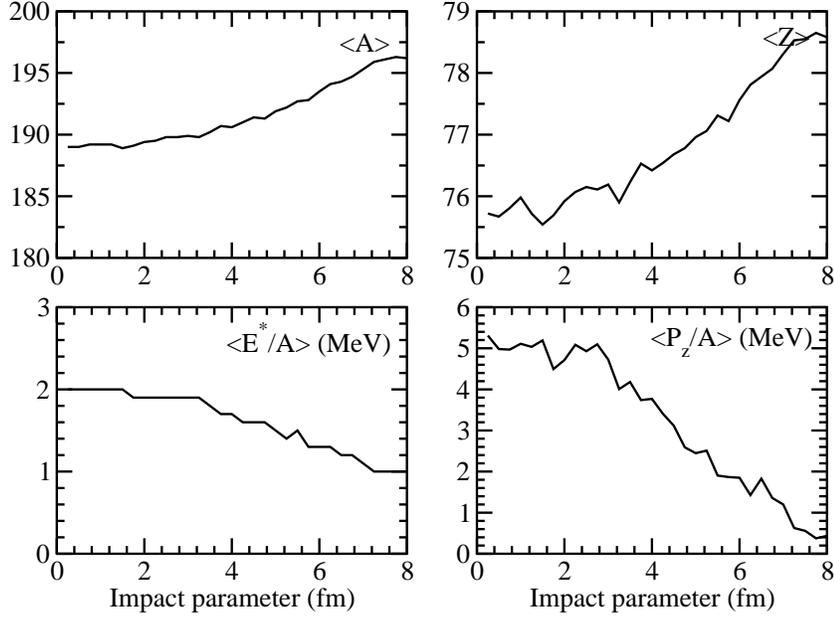}}}
\end{picture}
\caption{Impact parameter dependence of the mass (top-left), 
the charge (top-right panel), excitation energy per nucleon 
(bottom-left panel) and longitudinal momentum transfer per nucleon 
(bottom-right panel) of the fragmenting sources produced in 
p+Au reactions at $E_{lab}=0.8 GeV$.
}
\label{Fig1a}
\end{figure}
%%%%%%%%%%%%%%%%%%%%%%%%%%%%%%%%%%%%%%%%%%%%%%%%%%%%%%%%%%%%%%%%%%%
It turns 
out that for $p+Au$ reactions at $E_{beam}=0.8~AGeV$ incident energy the 
system approaches local equilibrium at $t\in [100,~120]~fm/c$, depending on the 
centrality of the proton-nucleus collision. During the dynamical evolution of 
a proton-nucleus collision in the spirit of Eq. (\ref{rbuu}) the nucleus gets excited 
due to momentum transfer and starts to emit nucleons (pre-equilibrium emission). 
Assuming that all particles inside the nuclear radius (including its surface) 
belong to the compound system, it is appropriate to define 
a {\it (fragmenting) source} by a density constraint of 
$\rho_{cut}=\frac{\rho_{sat}}{100}$ ($\rho_{sat}=0.16~fm^{-3}$ being 
the saturation density). We have checked that the results do not depend 
on the choice of the density constraint and the resulting difference is 
less than the statistical fluctuations. 
Thus, the parameters of the fragmenting source are given by the mass ($A$), 
the charge ($Z$) and the excitation energy at the time of equilibration. 

The major parameter entering into the SMM-code is the excitation 
energy $E_{exc}$ of a source with a given mass (A) 
and charge (Z) number. Its excitation is obtained 
by subtracting from the total energy the energy of the ground state, extracted 
within the same mean-field model as used in the Vlasov equation, for 
consistency. In a wide time interval from $t\approx 50~fm/c$ up to 
$t_{max}=150~fm/c$, in which one switches from dynamical to the statistical 
picture, the binding energy per particle of a ground state nucleus remains 
rather stable with small fluctuations around a mean value in the order 
of maximal $\pm 1\%$ using $200$ test particles per nucleon, which is 
important in calculating the excitation of the system.

All the theoretical results of the following section have been performed 
within the {\it hybrid} approach $GiBUU+SMM$ outlined here. Mass and charge 
numbers and excitation energy of the fragmenting source have been determined 
by imposing the density cut after onset of equilibration.

%%%%%%%%%%%%%%%%%%%%%%%%%%%%%%%%%%%%%%%%%%%%%%%%%%%%%%%%%%%%%%%%%%%
\section{Results on $0.8~GeV~~p+Au$ reactions}
%%%%%%%%%%%%%%%%%%%%%%%%%%%%%%%%%%%%%%%%%%%%%%%%%%%%%%%%%%%%%%%%%%%

%%%%%%%%%%%%%%%%%%%%%%%%%%%%%%%%%%%%%%%%%%%%%%%%%%%%%%%%%%%%%%%%%%%
\begin{figure}[t]
\unitlength1cm
\begin{picture}(10.,8.0)
\put(1.5,0.){\makebox{\psfig{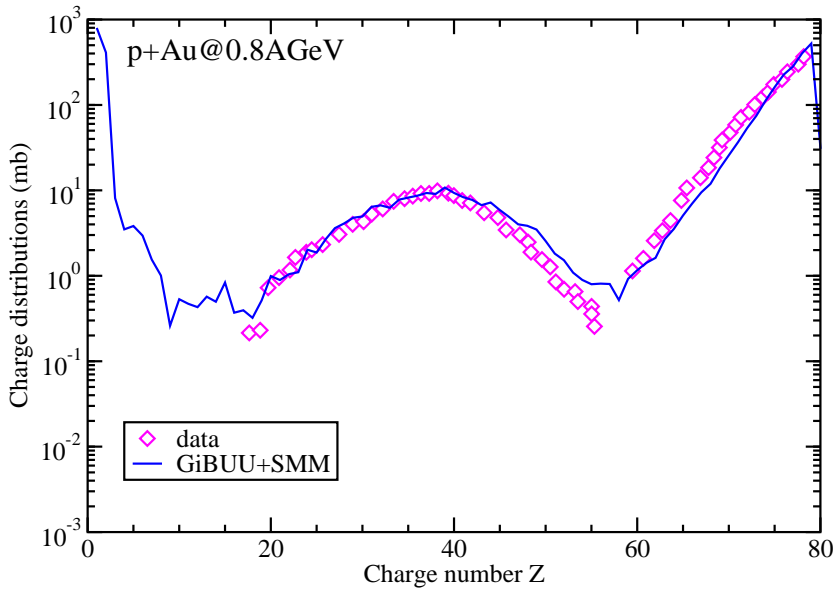}}}
\end{picture}
\caption{Charge distributions for $p+Au$ reactions at 
$E_{beam}=0.8~AGeV$ incident energy. Theoretical calculations 
(solid) are compared with data from \protect\cite{pAu2,pAu3} (open diamonds).
}
\label{Fig1}
\end{figure}
%%%%%%%%%%%%%%%%%%%%%%%%%%%%%%%%%%%%%%%%%%%%%%%%%%%%%%%%%%%%%%%%%%%
As a first benchmark we consider the properties of the initial non-equilibrium 
dynamics and the properties of the fragmenting source, see Fig.~\ref{Fig1a}.  
During the non-equilibrium dynamics the proton beam collides with nucleons 
of the target nucleus. The amount of energy transfer and thus of excitation of  
the residual nucleus with associated particle emission depends on the centrality 
of the reaction, as shown in Fig.~\ref{Fig1a}. With increasing impact 
parameter the proton beam experiences less collisions (and also less 
secondary scattering with associated inelastic processes, e.g. resonance 
production and absorption) with the particles of the target leading to 
less energy and momentum transfer. Thus, the pre-equilibrium emission 
is reduced, as can be clearly seen in Fig.~\ref{Fig1a}. In 
average, the pre-equilibrium emission takes mainly place in a time interval, 
in which the proton beam penetrates the nucleus. The time interval of pre-equilibrium 
emission only moderately depends on the impact parameter, in agreement with previous studies, 
e.g. see Ref.~\cite{cugnon}. However, as discussed in the previous section, we stop 
the dynamical calculation at a time later on, when all resonances have decayed and the 
residual system has achieved local equilibrium. The average amount of pre-equilibrium 
emission is $<A_{loss}>\approx 5$, $<Z_{loss}>\approx 3$ in 
terms of the mass and charge numbers, respectively. During this time interval the 
excitation of the residual nucleus drops from $<E_{exc}/A>\approx 4.2~MeV$ to 
$<E_{exc}/A>\approx (1.5-1.7)~MeV$ due to fast particle emission, before the 
residual system approaches a stable configuration. We note that our results are in 
agreement with those of other groups using non-relativistic approaches \cite{anna}, 
as expected and discussed above.

The hybrid model discussed in the previous section has been applied to 
$p+{}^{197}Au$ reactions at low relativistic energies, 
where a variety of experimental data is available \cite{pAu2,pAu3}. 
We have used 200 test particles per nucleon for each run and for 
each impact parameter from $b=0~fm$ up to $b_{max}$. 

\begin{figure}[t]
\unitlength1cm
\begin{picture}(10.,8.0)
\put(1.5,0.){\makebox{\psfig{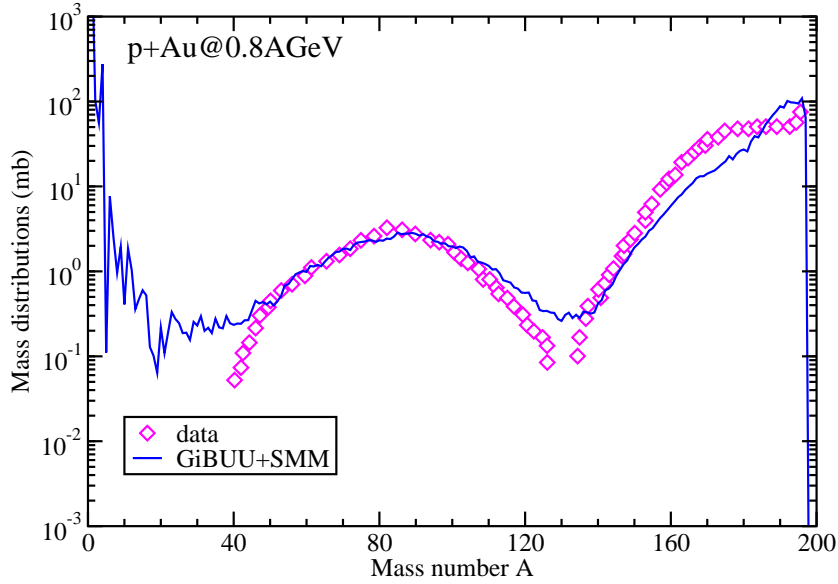}}}
\end{picture}
\caption{Same as in Fig.~\protect\ref{Fig1} for the mass distributions.
}
\label{Fig2}
\end{figure}

In Figs.~\ref{Fig1},~\ref{Fig2} we compare our theoretical results to 
experimental data \cite{pAu2,pAu3} for $p+Au$ reactions in terms of charge and mass 
distributions, respectively. The theoretical results are in reasonable 
agreement both with the {\it absolute} yields and the shape of the experimental 
data. Similar results were obtained in previous studies within the the INC model 
\cite{pAu1}.

%%%%%%%%%%%%%%%%%%%%%%%%%%%%%%%%%%%%%%%%%%%%%%%%%%%%%%%%%%%%%%%%%%%
%%%%%%%%%%%%%%%%%%%%%%%%%%%%%%%%%%%%%%%%%%%%%%%%%%%%%%%%%%%%%%%%%%%
The fragmentation of an excited system is a complex process involving different 
mechanisms of dissociation: Sharp peaks at the regions 
$(A,Z)=(A,Z)_{init}$ and $(A,Z)\approx (A,Z)_{init}/2$, where 
$A_{init},~Z_{init}$ denote the initial mass and charge numbers, respectively, 
correspond only to the most peripheral events with very low momentum 
transfer and thus with low excitation energy. According to the SMM model, 
heavy nuclei at low excitation energy corresponding to a temperature $T<2~MeV$ 
mainly undergo evaporation and fission producing the sharp peaks at the 
very high and low mass and charge numbers. With decreasing centrality the 
excitation energy, respectively temperature, increases. As the excitation 
energy (or temperature) approaches $T\approx 5~MeV$ the sharp 
structure degrades due to the onset of the multifragmentation mechanism, and at 
higher excitations, $T=5-17~MeV$, one expects exponentially decreasing yields 
with decreasing mass/charge number. These different phenomena of dissociation 
of an excited source finally leads to wide distributions in $A,~Z$ as shown in 
Figs.~\ref{Fig1},~\ref{Fig2}. 

The main features of the fragmentation process are apparently reasonably well 
reproduced by our hybrid approach. The combination of 
the non-equilibrium dynamics (GiBUU) and the statistical decay of the 
equilibrated configuration (SMM) obviously accounts for the essential aspects 
of the reaction. The first stage is important in calculating 
the excitation of the residual (due to dynamical pre-equilibrium emission) system, 
and the second one for the statistical fragmentation of the equilibrated 
configuration.

%%%%%%%%%%%%%%%%%%%%%%%%%%%%%%%%%%%%%%%%%%%%%%%%%%%%%%%%%%%%%%%%%%%
\begin{figure}[t]
\unitlength1cm
\begin{picture}(10.,8.0)
\put(1.5,0.){\makebox{\psfig{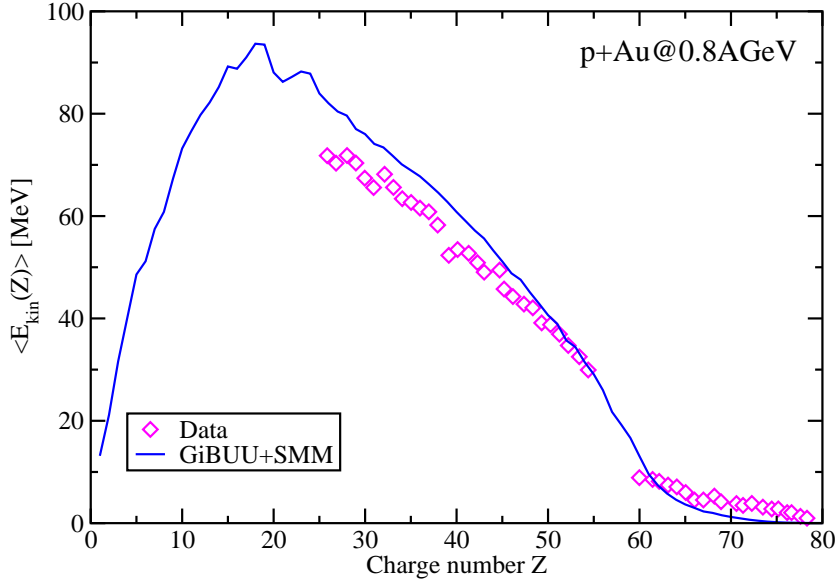}}}
\end{picture}
\caption{Same as in Fig.~\protect\ref{Fig1} for the average kinetic energies.
}
\label{Fig3}
\end{figure}
%%%%%%%%%%%%%%%%%%%%%%%%%%%%%%%%%%%%%%%%%%%%%%%%%%%%%%%%%%%%%%%%%%%

The situation is quite similar for the mean kinetic energies of 
produced fragments, displayed in Fig.~\ref{Fig3}. The average kinetic energies 
show an almost linear rise from the low-$Z$ and high-$Z$ regions towards 
the maximum around $Z\approx 20$. Qualitatively, the lightest fragments 
with the larger slopes are produced in the most central collisions, 
corresponding to a large momentum transfer. The low energy tail reflects 
the most peripheral events with low momentum transfer. It mainly contains the 
heavy residual products with $Z>60$. 

An exact interpretation of the system size dependence of the average 
kinetic energies in Fig.~\ref{Fig3} is not trivial and would require a 
detailed discussion of the SMM model, which is not the scope of this work. 
It is possible, however, to give some quantitative arguments. 
Quantitatively, for a pure binary fission one would expect naively a 
maximum of the energy distribution at about half the target charge, 
$Z=40$. However, such a picture would neglect the dynamics of the 
reaction. In order to understand the energy distribution we have to 
consider the BUU pre-equilibrium dynamics and the fragment formation 
mechanism. The transport dynamics leads to pre-equilibrium particle 
emission which results to an initial compound nucleus with smaller 
$Z$-values compared to that of the target. Thus the energy naturally 
becomes smaller. On the other hand, 
the formation of the fragments is determined by the nuclear binding 
energies, and their interaction after formation where Coulomb effects 
play an important role \cite{SMM}. The binding energy effect favors 
nuclei around iron ($Z\sim 26$). The Coulomb repulsion leads to 
long-range correlations among the fragments with a tendency to 
shift the distribution to slightly smaller $Z$-values. 

\begin{figure}[t]
\unitlength1cm
\begin{picture}(10.,10.0)
\put(0.5,0.){\makebox{\psfig{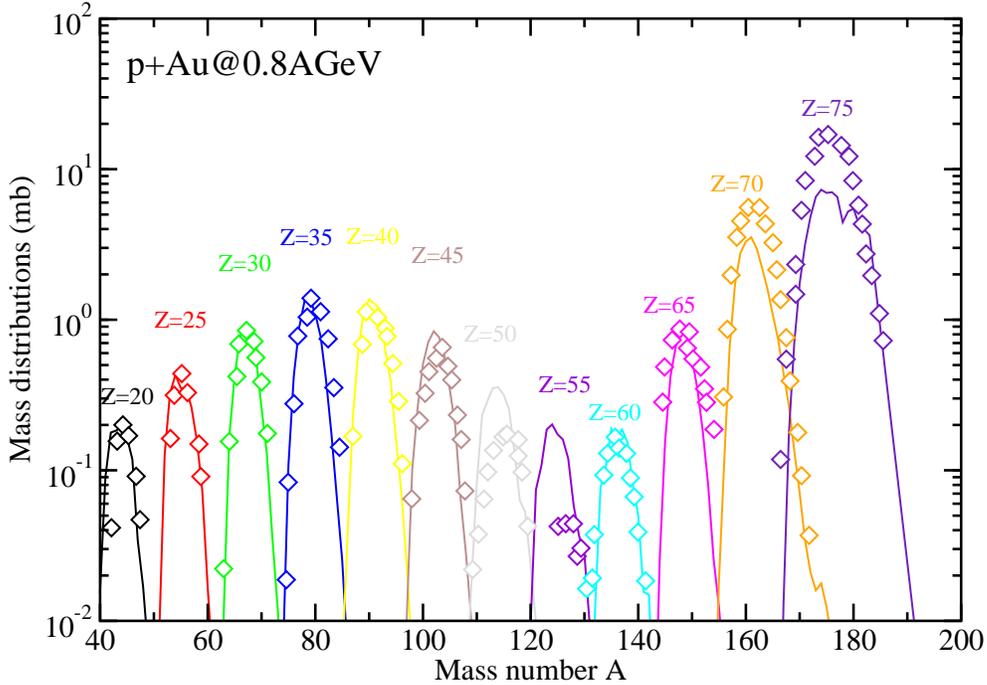}}}
\end{picture}
\caption{Same as in Fig.~\protect\ref{Fig1} for the 
mass distributions of different isotopes (indicated by the 
atomic number $Z$).
}
\label{Fig4}
\end{figure}

\begin{figure}[t]
\unitlength1cm
\begin{picture}(10.,10.0)
\put(0.5,0.){\makebox{\psfig{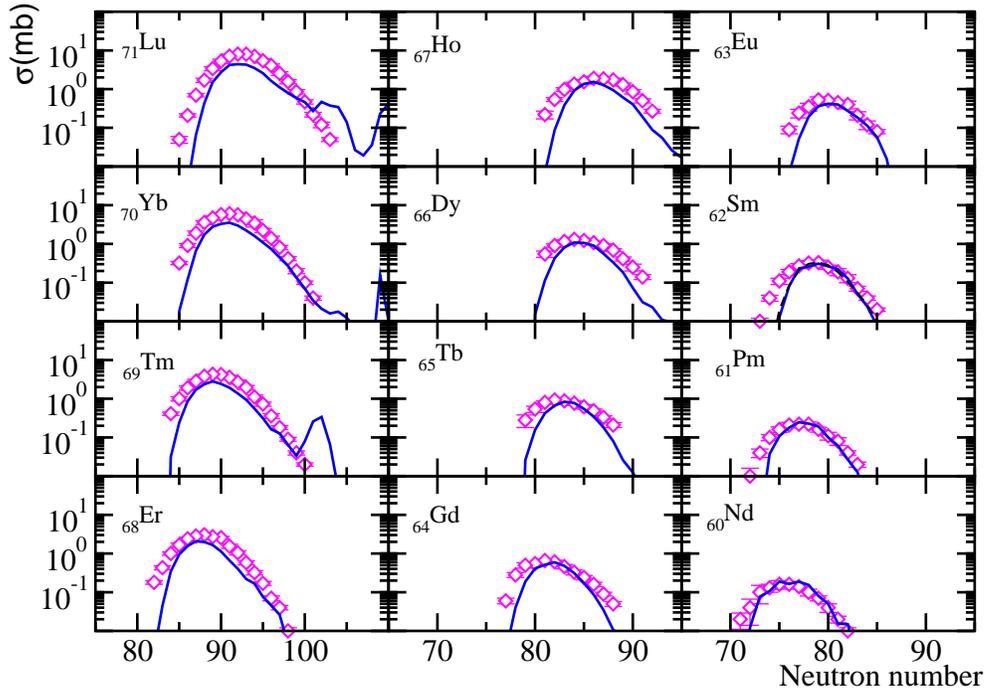}}}
\end{picture}
\caption{Same as in Fig.~\protect\ref{Fig1} for the 
neutron distributions of different isotopes (indicated by the 
atomic number $Z$). The meaning of the curves is the same as in 
Fig.~\protect\ref{Fig1}, expect that the experimental data are 
taken from \protect\cite{pAu2,pAu3}.
}
\label{Fig5}
\end{figure}

The mass and charge distributions of the yields or energies of the 
produced fragments show only the general trends, which are reproduced 
well by the theoretical model applied in this work. However, 
they are not sensitive enough to the details of the reaction. A more 
stringent test is to study the characteristics of individual nuclides and 
particles produced in the reaction. Figs.~\ref{Fig4} and 
\ref{Fig5} show theoretical results and experimental data on production 
yields of different separate isotopes in the atomic and mass number, 
respectively. A good overall agreement is achieved, again 
as a good check of the theoretical hybrid model. 

In particular, in Fig.~\ref{Fig4} we see that for isotopes 
produced in the spallation region (not too far from the target mass) and for 
fission fragments not too far from the maximum yield the comparison 
between the theory and experiment is only satisfactory. Similar trends 
are observed in the neutron spectrum of separate isotopes, see again 
Fig.~\ref{Fig5}. Interesting is the discrepancy in the proton-rich 
regions of the isotopic distributions of the heaviest elements, which 
can be also seen in the global $Z$- and $A$-distributions in the 
corresponding region. 
%%%%%%%%%%%%%%%%%%%%%%%%%%%%%%%%%%%%%%%%%%%%%%%%%%%%%%%%%%%%%%%%%%%
%%%%%%%%%%%%%%%%%%%%%%%%%%%%%%%%%%%%%%%%%%%%%%%%%%%%%%%%%%%%%%%%%%%
This detailed comparison shows the limitations 
and needed improvements of the hybrid model. The pre-equilibrium dynamics 
seems to lead to a more excited configuration with respect to the 
experiment, since all the theoretical distributions are moderately smoother in 
comparison with the experimental data, which is better visible in the 
detailed isotopic distributions. 

In general, it turns out that the hybrid model gives a quite satisfactory
description of fragmentation data, which is a non-trivial task in transport 
dynamical approaches. We note again, that the non-equilibrium dynamics has been 
treated in a microscopic way using the relativistic coupled-channel transport 
approach, which is an important step in extracting the properties of the fireball-like 
configuration, before applying its statistical decay in the spirit of the 
SMM model. 

%%%%%%%%%%%%%%%%%%%%%%%%%%%%%%%%%%%%%%%%%%%%%%%%%%%%%%%%%%%%%%%%%%%
\section{Conclusions and outlook}
%%%%%%%%%%%%%%%%%%%%%%%%%%%%%%%%%%%%%%%%%%%%%%%%%%%%%%%%%%%%%%%%%%%

%%%%%%%%%%%%%%%%%%%%%%%%%%%%%%%%%%%%%%%%%%%%%%%%%%%%%%%%%%%%%%%%%%%
%%%%%%%%%%%%%%%%%%%%%%%%%%%%%%%%%%%%%%%%%%%%%%%%%%%%%%%%%%%%%%%%%%%
We have investigated the fragmentation mechanism within a hybrid approach 
consisting of a dynamical transport model of a BUU type and a statistical 
one in the spirit of the Statistical Multifragmentation Model (SMM), and applied 
it to low energy proton-induced reactions by means of fragment formation.

The main contribution was to show the reliability and possible limitations 
of the dynamical transport model for the description of 
multifragmentation within additional statistical approaches. In particular, 
it turned out that the hybrid model describes a wide selection of 
experimental data reasonably well. 

As a future project, a consistent description of the 
initial ground state might be achieved within a semi-classical 
density functional theory by determining the energy density functional 
consistently with the density profile of a nucleus implying also the 
inclusion of surface and isovector contributions to the energy density and 
thus to the dynamical evolution. This work is currently in progress.

It is worthwhile to note, that the GiBUU transport approach contains 
the production and propagation of baryons and mesons with strangeness in 
appropriate relativistic mean fields. Also the SMM code has been extended to statistical 
decay of fragments with finite strangeness content ({\it hypernuclei}) 
\cite{botvinaHyp}. Therefore, motivated by the results of this work it is 
straightforward to continue this field of research investigating hypernucleus 
production from highly energetic nucleus-nucleus collisions. This part of study 
is still in progress.

In summary, we conclude that this work provides an appropriate theoretical basis 
for investigations on fragmentation with a new perspective for hypernuclear physics.

{\it Acknowledgments.}  We would like to acknowledge Prof. A. Botvina for many 
useful discussions and for providing us with the SMM-code. We also thank 
the GiBUU group for many useful discussions. This work is 
supported by BMBF.

%\section*{References}
%%%%%%%%%%%%%%%%%%%%%%%%%%%%%%%%%%%%%%%%%%%%%%%%%%%%%%%%%%%%%%%%%%%%%%%%%
%                                                                       %
%   BEGIN OF BIBLIOGRAPHY                                               %
%                                                                       %
%%%%%%%%%%%%%%%%%%%%%%%%%%%%%%%%%%%%%%%%%%%%%%%%%%%%%%%%%%%%%%%%%%%%%%%%%

\end{document}